\documentclass[twocolumn,trackchanges]{aastex631}
\usepackage{lineno}
\usepackage{natbib}

\shorttitle{Tidal Forces on Molecular Clouds}
\shortauthors{Lee, Saur, Mac Low, \& Li}

\graphicspath{{./}{figures/}}

\begin{document}
\title{The Importance of Tidal Forces in Molecular Cloud Dynamics}

\correspondingauthor{Hui Li}
\email{hliastro@tsinghua.edu.cn}

\author[0009-0000-6106-9552]{JinWoo Lee}
\altaffiliation{Co-first author}
\affiliation{Department of Astronomy, Columbia University, New York, NY 10025, USA}
\affiliation{Department of Physics and Astronomy, Seoul National University, Seoul 08826, Republic of Korea}

\author[0009-0008-1361-9458]{Alexa Saur}
\altaffiliation{Co-first author}
\affiliation{Department of Computer Science, Worcester Polytechnic Institute, Worcester, MA}

\author[0000-0003-0064-4060]{Mordecai-Mark Mac Low}
\affiliation{Department of Astrophysics, American Museum of Natural History, New York, NY 10024, USA}
\affiliation{Department of Astronomy, Columbia University, New York, NY 10025, USA}

\author[0000-0002-1253-2763]{Hui Li}
\affiliation{Department of Astronomy, Tsinghua University, Beijing 100084, People’s Republic of China}

\begin{abstract}

We investigate the role of tidal forces in molecular cloud formation by examining how apparent boundedness, as diagnosed by the classical virial parameter, relates to the actual gravitational state of clouds subject to tidal forces from their environment. Clouds are identified by a dendrogram algorithm in zoom-in regions taken from a simulation of a Milky Way-mass galaxy with the Voronoi mesh code AREPO that resolves star-forming regions at sub-parsec resolution. To look at a range of environments, we use data from three different regions that evolve differently in the center, near the equivalent of the Solar circle, and the outskirts of the modeled galaxy, at three different times, each spaced 2 Myr apart. We compute the importance of tidal forces on all identified clouds. We then compare the boundedness of clouds including only their internal potentials to boundedness also including the external gravitational potential. This comparison shows that tidal forces can unbind apparently bound clouds and bind apparently unbound clouds. We characterize the cloud population by comparing their virial parameters to their surface densities, finding the ratio of the maximum to the minimum eigenvalues of the tidal tensor, and determining the strength of gravitational instability in each examined region. We find that it is necessary to take the total gravitational potential into account rather than just the internal self-gravity of the clouds to have an accurate understanding of cloud dynamics. 

\end{abstract}

\section{Introduction} \label{sec:intro}

Molecular clouds are the coldest and densest constituents of the interstellar medium, which have long been identified as the birth sites of young stars in galaxies \citep{shu_etal1987,heyer_dame2015}. Yet, when it comes to understanding the dynamics of these clouds, gaps remain because most models neglect the gravitational influence of the environment on forming clouds. 
Determining the relative importance of the external and internal gravitational potential is required to understand how important the environment is in determining whether stars can form in a cloud.

Molecular clouds satisfy the time-dependent virial equation \citep{mckee1992}. The usual approximation used by observers to assess whether a molecular cloud is bound by gravity or dominated by turbulence is to ignore the time-dependent change in its moment of inertia, as well as all surface terms and compute what we will call the observational virial parameter, which gives the virial ratio for an equivalent isolated homogeneous sphere \citep{bertoldi1992},
\begin{equation}\label{alpha_class_eq}
    \alpha_{\rm obs} = \frac{5 \sigma_{\rm 1D}^2 R}{GM},
\end{equation}
where $R$ is the equivalent radius of the cloud, usually derived from its area, $\sigma_{\rm 1D}$ is the one-dimensional velocity dispersion along the line of sight, $M$ is the total cloud mass, and $G$ is the gravitational constant.  If $\alpha_{\rm obs} < 1$, the cloud is argued to be bound.

However, \citet{galeano} suggest that to determine whether clouds are bound or not, it is important to compare the kinetic energy of the cloud with the full potential energy, including external extensive or compressive tidal forces produced by nearby clouds or the spiral arms, disk, or bulge of the galaxy.
We compare results from determining whether clouds are bound by using the classical virial parameter including only the mass of the cloud itself to a determination accounting for the full gravitational potentials of clouds, including the tidal field produced by their environment.

We analyze a population of clouds drawn from a simulation of a Milky Way-mass galaxy using a novel zoom-in technique that allows target star-forming regions to be well-resolved at sub-parsec resolution. The zoom-in starts from a global galaxy simulation \citep{hui2} that uses SMUGGLE \citep{smuggle}, a comprehensive interstellar medium and stellar feedback model for the moving-mesh code AREPO \citep{springel2010}.

We examine three different regions of the simulated galaxy---central, intermediate, and outskirts---to test the role of tidal forces in different locations within a galaxy. We  characterize the cloud population by examining the \citet{heyer} relation between the virial parameter and the column density of the clouds, as well as by determining the ratio of maximum to minimum eigenvalues of the tidal tensor including the environment. 

We study whether differences arise between the different regions, including whether the center of the galaxy will be comparatively less efficient in forming stars due to tidal disruption. By studying the internal and external potentials in these regions across the three different time epochs, we determine the role of tidal forces in the energy budget of the clouds.

In Sect.\ \ref{sec:methods}, we discuss our analysis of the simulation data, focusing on evaluation of the tides acting on the gas clouds and characterization of their physical properties. 
In Sect.\ \ref{sec:results}, we present and analyze the results while suggesting the underlying physical properties that might explain them. Lastly, in Sect.\ \ref{sec:conclusions}, we conclude with a summary and discussion.

\section{Methods} \label{sec:methods}
\subsection{Simulations}
Our analysis uses data from a set of hydrodynamical simulations of a Milky Way-mass galaxy, performed with a novel zoom-in technique that allows target star-forming regions to be well-resolved at sub-parsec scale (Li et al.\ in prep.). The simulations were performed using the Voronoi unstructured mesh hydrodynamical code AREPO \citep{springel2010}. The zoom-in starts from the fiducial run SFE1 in \citet{hui2} using the SMUGGLE framework \citep{smuggle}. This framework incorporates gas cooling and heating over a temperature range of 10--$10^8$ K so that the thermodynamic properties of the multiphase ISM are modeled self-consistently. Star formation and stellar feedback processes including ionizing radiation, winds, and supernovae are implemented with subgrid models. The initial conditions of the isolated disk galaxy consist of a live stellar bulge and disk, a gaseous disk, and a fixed \citet{navarro1997} dark matter halo, all having masses similar to the Milky Way's components. The galaxy develops a flocculent disk without a central bar.

Twenty four star formation complexes of interest at different galactic locations were selected and traced via Monte Carlo tracer particles so that the time evolution of their Lagrangian regions are tracked accurately. We then rewind the simulations and adaptively and gradually refine the Lagrangian regions of the complexes to a mass resolution of $1 M_\odot$. The zoom-in region are run together with the background galaxy, so the whole simulation suite achieves a huge dynamical range from the $200$ kpc halo scale down to the $\sim0.1$ pc molecular core scale. This zoom-in suite provides the necessary data to compute the kinetic and gravitational energies, virial parameters, and tidal acceleration tensors needed to understand the dynamics of the clouds.

To simplify the analysis, we select three specific zoom-in regions, all 1 kpc on each side, which represent three different ISM environments: 1) the {\em central region} whose center of mass has a  galactocentric radius of 0.27 kpc; 2) the {\em intermediate region} whose center of mass has a galactocentric radius of 8.4 kpc, similar to the solar neighborhood; and 3) the {\em outskirt region} whose center of mass has a galactocentric radius of 12 kpc. These three regions cover a wide range of galactocentric radius and ISM conditions.

For our analysis, we identified clouds across the three zoom-in regions with data from three epochs for each region, beginning 50 Myr after the start of the larger simulation and spaced 2 Myr apart. These epochs, numbered 520, 540, and 560, were selected to fall during the main star formation episode in this isolated galaxy. Because of the lack of continuing gas infall, star formation does not continue in an equilibrium state thereafter, but rather dies away.

\begin{figure*}
\centering
\includegraphics[width=0.9\linewidth]{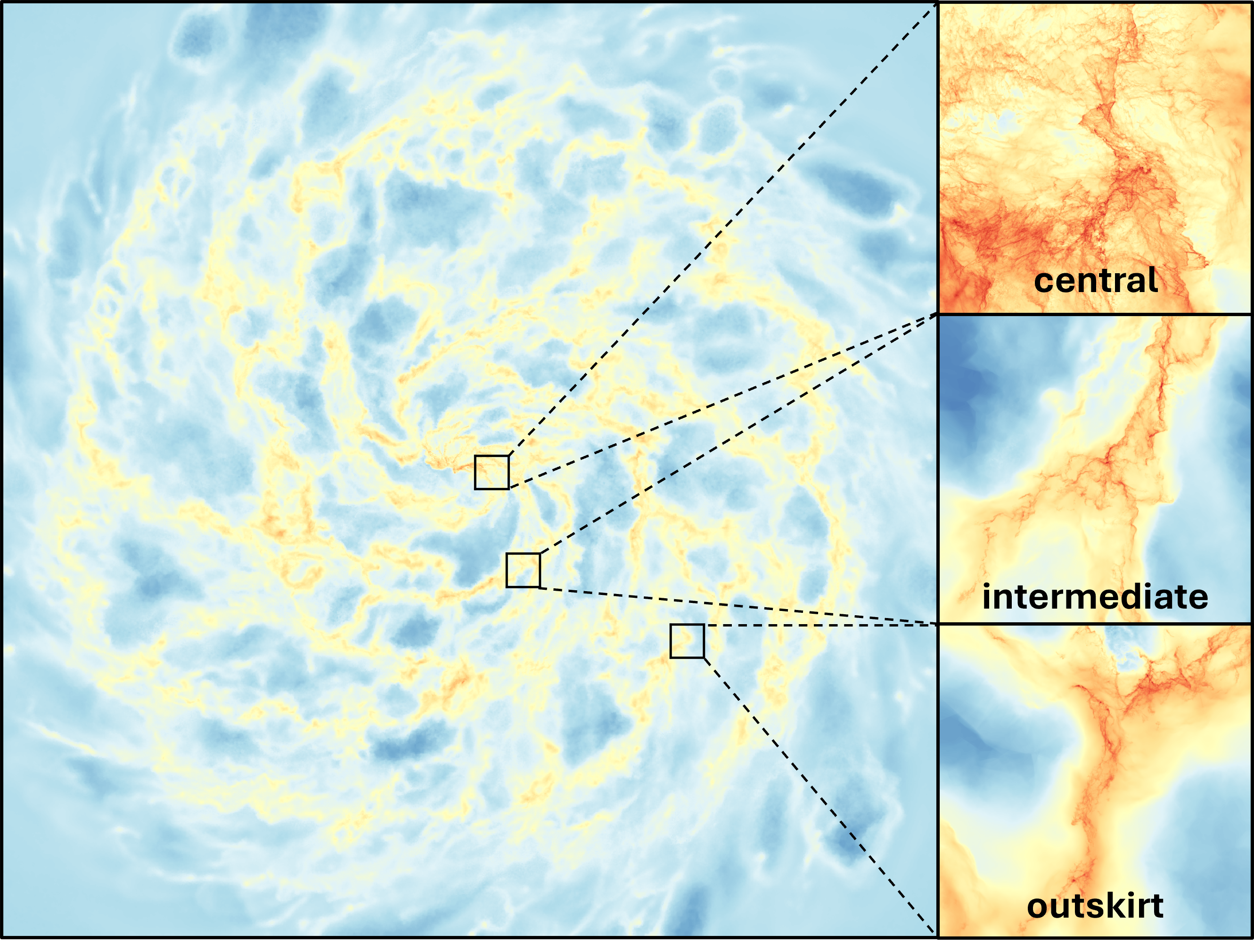}
\caption{\label{prj} Gas density projection plot of the whole galaxy (left) and the three zoom-in regions: central (upper right), intermediate (mid right), and outskirt (lower right). The three zoom-in figures show gas surface density in the range of $10-10^6 M_\odot/{\rm pc}^2$.}
\end{figure*}

\subsection{Cloud Analysis}

Each simulation epoch examined was converted from the original irregular Voronoi mesh to a $1000^3$ Cartesian grid with 1~pc zone size. In order to convert the data, the particle masses, velocities, densities, and internal energies were merged through a sum, mass-weighted average, volume-weighted average, and mass-weighted sum, respectively. Each particle has a Voronoi mesh volume, a mass, a velocity, and an internal energy associated with it. Other variables were calculated from these quantities.

\begin{figure*}
\centering
\includegraphics[width=0.9\linewidth]{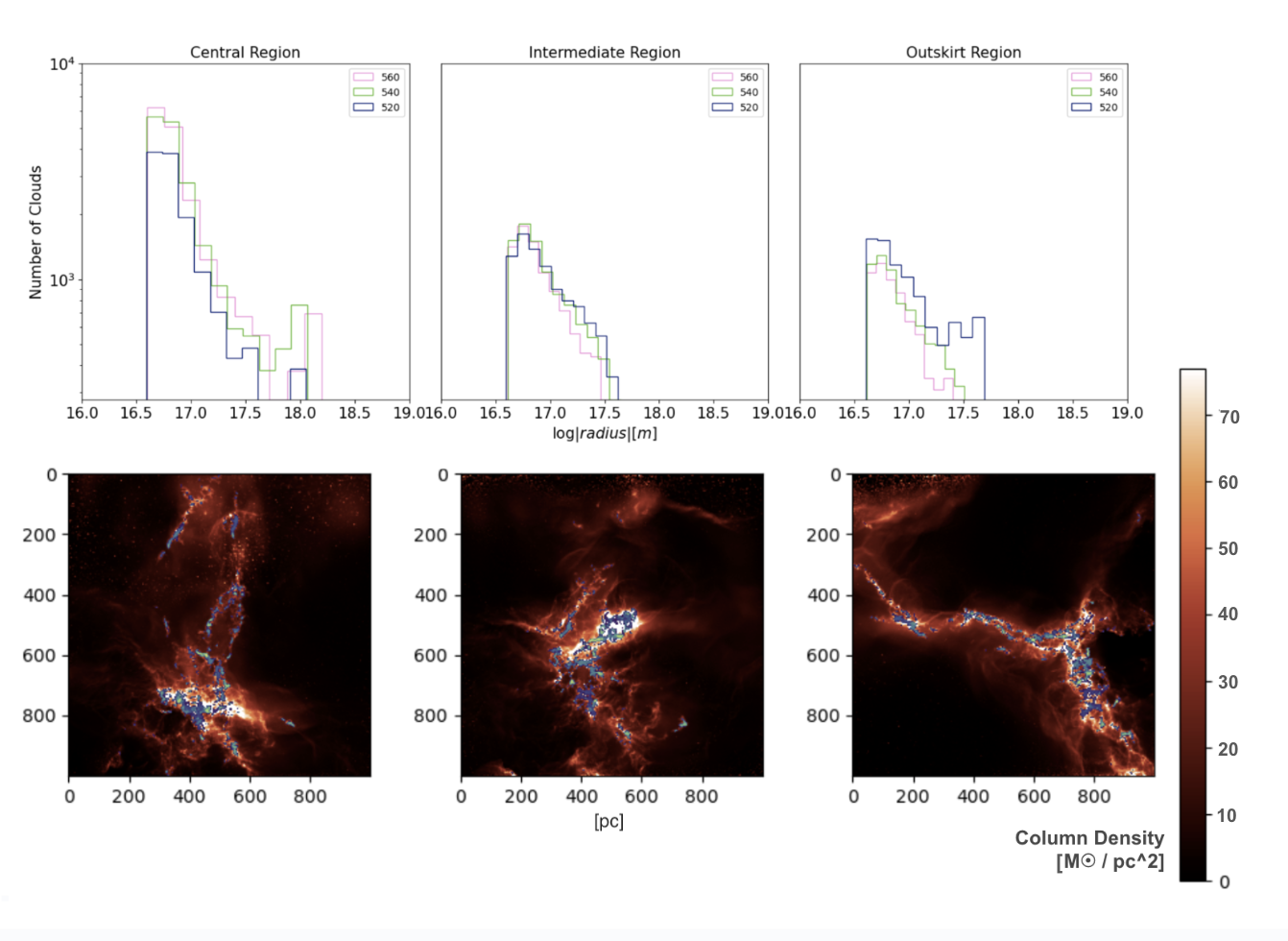}
\caption{\label{snapshot} {\em (Top)} Size distribution in meters of all clouds across three different regions at three epochs.  The minimum cloud radius is derived from our choice of {\tt minnpix} $=10$~voxels. {\em (Bottom)} Column densities in three different angles of epoch 540 in the intermediate region only. The three panels for the bottom figure from left to right each correspond to projections along the x-, y-, and z-axis, respectively. Overlapping in green-purple colors are the clouds making up the dendrogram.}
\end{figure*}

We identified clouds in each epoch using the dendrogram algorithm implemented in the Python package {\sc astrodendro} \citep{dendrograms} on the Cartesian grid. There are three parameters that control the {\sc astrodendro} algorithm for determining when density perturbations should be treated as independent nodes or leaves: the threshold value of density {\tt minvalue}; how the strength of the contrast to the surroundings {\tt mindelta}; and the minimum number of voxels in the structure {\tt minnpix}. We varied these parameters to extract a cloud population with mass, density, and length scales characteristic of Milky Way clouds. Our chosen parameter values are {\tt minvalue} $= 0.5$, {\tt mindelta} $=0.5$, in code density units of 10$^{10}M_\odot$ kpc$^{-3}$, and {\tt minnpix} $=10$~voxels. (Note that 1 code density unit $\simeq 300$~cm$^{-3}$ for neutral gas with 10\% He by number). {\sc astrodendro} provides a tree containing many substructures for each epoch, and we include all structures in our calculations and plots. The number of clouds varies between epochs and regions. We have shown the column density of epoch 540 in the intermediate region as an example in Figure~\ref{snapshot}.

The radius of each cloud was estimated as the radius of the equivalent sphere:
\begin{equation}\label{radius_eq}
    R = \left(\frac{3}{4\pi}V\right)^{1/3},
\end{equation}
where $V$ is the sum of the volumes of the cells belonging to the cloud.

We compute two different virial parameters in order to understand the underlying differences between the dynamics of an isolated, homogeneous cloud (Eq.~\ref{alpha_class_eq}) and those of a cloud taking into account the total gravitational potential energy of its environment. Because we did not make true simulated observations, instead of using Equation~(\ref{alpha_class_eq}), we followed the classical approximation to the virial parameter \citep[e.g.][]{bertoldi1992}
\begin{equation}\label{alpha_class}
   \alpha_{\rm class}=\frac{2K}{\left| E_{g} \right|},
\end{equation}
where the gravitational energy of an isolated, homogeneous sphere
\begin{equation}\label{peclas}
    E_{g}=-\frac{3}{5}\frac{GM^{2}}{R}.
\end{equation}
The kinetic energy of each cloud is computed by summing the square of the velocity ${\mathbf v}_i$ of each cloud cell relative to the center of mass multiplied by its mass
\begin{equation}\label{kinetic_eq}
    K =  \frac{1}{2}\sum_i  m_i {\mathbf v}_i \cdot {\mathbf v}_i.
\end{equation}

In order to capture the full effect of gravity from internal and external sources acting on clouds, we compute
the total gravitational energy of the clouds \citep{shu,mckee1992,galeano}
\begin{equation}\label{work_eq}
    W=-\sum_{i}^{}x_{i}\rho_i \frac{\partial\Phi}{\partial x_{i}}\Delta V_i.
\end{equation}
In detail, for each cloud, to compute the total gravitational energy, the derivative of the cloud's gravitational potential $\phi_i$ was computed at every point within the cloud, using numerical differences between zones to take the
derivatives, and multiplied with the position vector ${x_i}$ and the cell volume $\Delta V_i$, then summed over the volume.

Taking both the kinetic and the potential energy of the cloud, the full virial parameter can be defined as
\begin{equation}\label{alpha_full}
    \alpha_{\rm full} = 2K/W.
\end{equation}
Note that in this case, the virial parameter can be either positive or negative depending on the sign of the potential energy W. A positive virial parameter $0 < \alpha_{\rm full} < 2$ reflects an object unbound primarily by tidal forces, while an object with $|\alpha_{\rm full}| > 2$ is unbound  primarily by kinetic energy \citep{galeano}.

\cite{ganguly} defined a virial parameter including both 
the turbulent kinetic energy $k$ and the thermal energy $E_{\rm TE}$
\begin{equation}\label{alpha_vol}
    \alpha_{\rm tt}=\frac{2K + 2E_{\rm TE}}{|W|},
\end{equation}
which is positive definite. Thermal energy $E_{\rm TE}$ was computed by summing the mass weighted internal energy of each point in the cloud.

We also computed the volume average of the tidal tensor
\begin{equation}\label{eigen_eq}
    \left \langle T \right \rangle_{ij} = \frac{1}{V} \int_{V}^{}\partial _{i}\partial _{j}\Phi_{ij}  d^3r,
\end{equation}
where $\Phi_{ij}$ is the gravitational potential at each point and we used finite differences between zones to take the derivatives. This tensor was then transformed to orthogonal form to compute its eigenvalues and analyze its behavior.
\begin{figure*}
\centering
\includegraphics[width=0.9\linewidth]{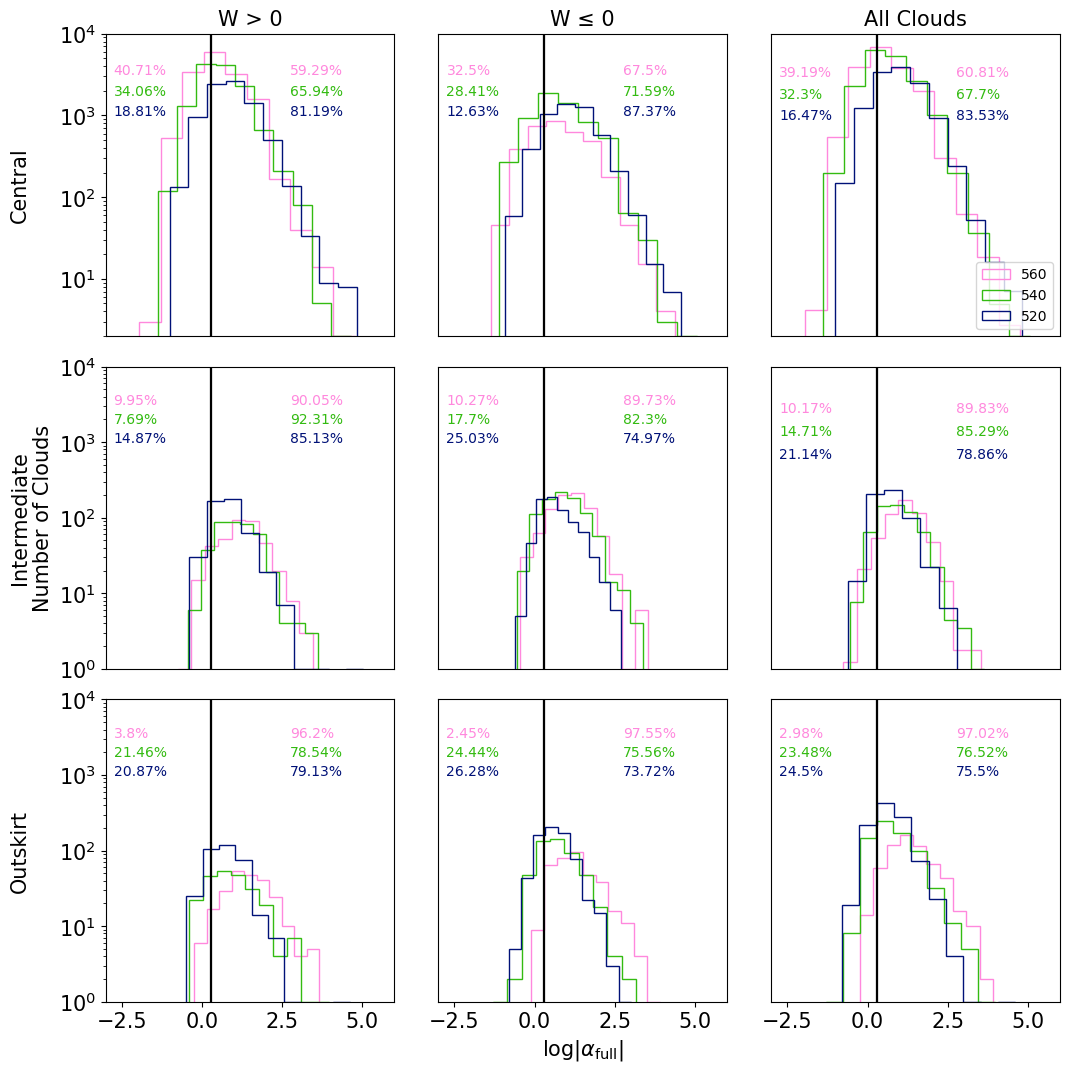}
\caption{\label{fig:alpha_hist} Histograms of the virial parameter $\alpha_{\rm full}$ for clouds that are unbound, with gravitational energy $W > 0$ {\em (left column)}, and either bound or unbound with $W \leq 0$ {\em (middle column)}, as well as for the full population {\em (right column)}. The three different epochs in each panel across the three regions are indicated by the legend. In the middle column, the vertical line at $|\alpha_{\rm full}| = 2$ 
marks the division between bound {\em (left)} and unbound {\em (right)} clouds according to the full virial parameter criterion. 
The percentages of clouds on either side of the critical virial parameter, color coded by epoch, are shown in the upper corners of each panel.
}
\end{figure*}

\begin{figure*}
    \centering
    \includegraphics[width=0.9\linewidth]{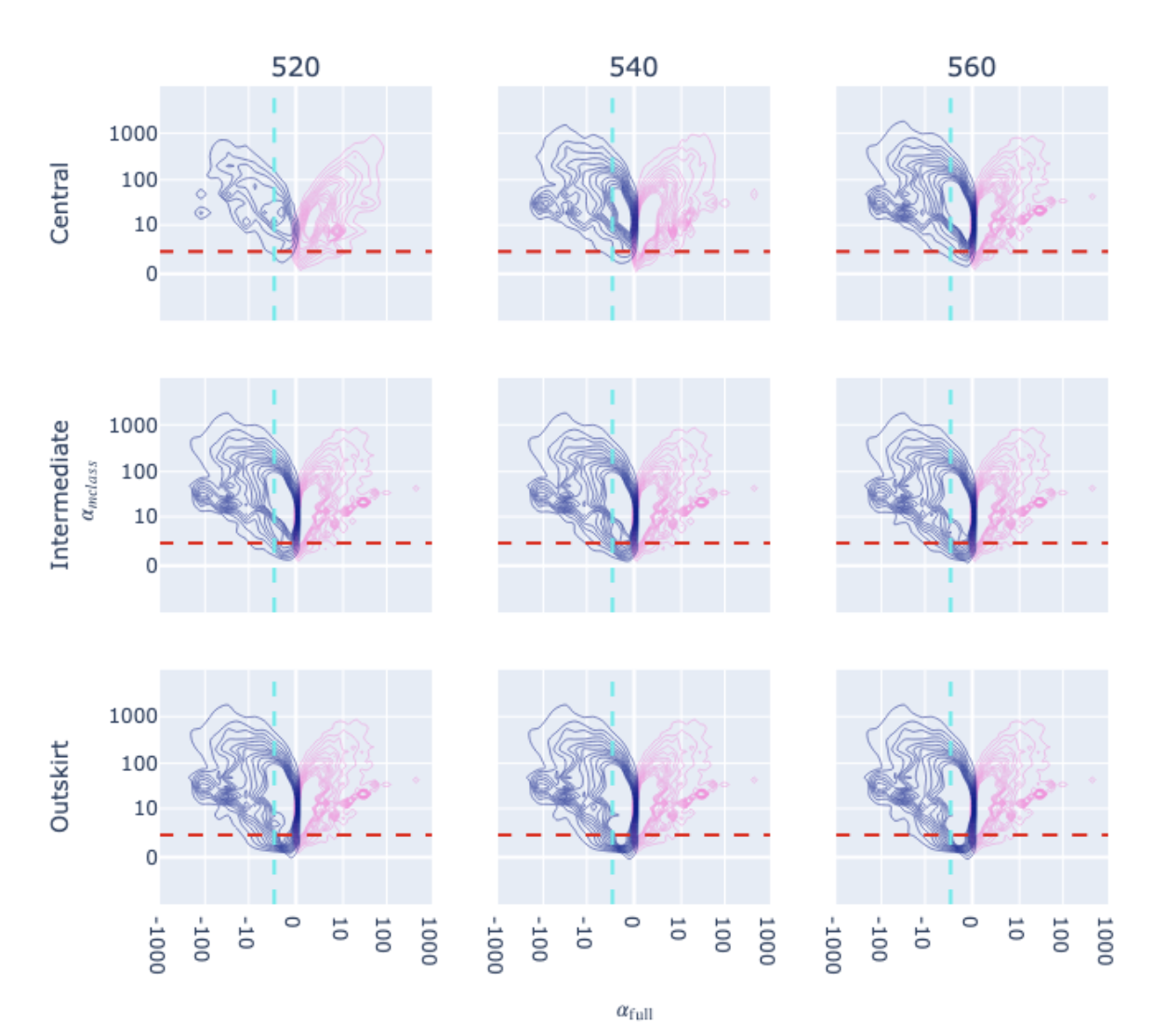}
    \caption{\label{fig:signed} Contour plots of the classical virial parameter $\alpha_{\rm class}$ against the signed value of the full virial parameter $\alpha_{\rm full}$. The {\em pink} contours contain clouds with $\alpha_{\rm full} > 0$, while the {\em dark blue} contours show $\alpha_{\rm full} < 0$. Clouds that appear bound with $\alpha_{\rm class} < 2$ (below the  {\em horizontal red dashed} line) can still have values of $ 0 < \alpha_{\rm full} < 2$ showing that they are unbound because of tidal forces, or $|\alpha_{\rm full}| > 2$ showing they are unbound because of turbulence. Clouds that appear unbound with $\alpha_{\rm class} > 2$ can still have  values of $0 > \alpha_{\rm full} > -2$ ({\em blue} contours right of {\em vertical blue dashed} line), showing that they are bound because of tidal forces. Columns show results from epochs 520, 540, and 560, while rows show central, intermediate and outskirt regions. 
    Note the nonuniform symmetric log scale on the horizontal axis, which becomes linear rather than logarithmic for $|\alpha| < 1$.}
\end{figure*}

Lastly, to determine the gravitational stability of each region, we calculated the Toomre gravitational instability parameter following \citet{meng2019} \citep[see also][]{wang1994,romeo2011}:
\begin{equation}\label{eq:qtot}
\frac{1}{Q} \simeq \frac{1}{Q_{\rm gas}} + \frac{1}{Q_{\rm star}},
\end{equation}
where the gravitational instability parameter for gas \citep{goldreich_lynden} 
\begin{equation}\label{q_eq}
    Q_{\rm gas}=\frac{\sigma_{\rm gas} \kappa }{\pi G \Sigma_{\rm gas}},
\end{equation}
and the parameter for the stars \citep{toomre1964}
\begin{equation}
     Q_{\rm star} = \frac{\sigma_* \kappa}{\pi G \Sigma_{\rm star}}.
\end{equation}
In calculating the Toomre parameter, the epicyclic frequencies $\kappa$ of the different regions in the disk were computed along with their total gas velocity dispersion 
\begin{equation}
\sigma_{\rm gas}=(\sigma_{\rm turb}^2+c_s^2)^{1/2}
\end{equation}that considers both the turbulent $\sigma_{\rm turb}$ and thermal $c_s$ components, stellar velocity dispersions $\sigma_*$ and surface densities $\Sigma_{\rm gas}$ and $\Sigma_{\rm star}$. 
For simplicity, we approximated $\kappa$ by a spherically symmetric expression in the limit of a flat rotation curve:
\begin{equation}
\kappa=\kappa_c \equiv \frac{2V_c^2(R)}{R^2},
\end{equation}
where $V_c(R)=(GM(R)/R)^{1/2}$ at galactocentric radius $R$, within which the total enclosed mass is $M(R)$.
For the kiloparsec-sized central zoom-in region,
the variation of $\kappa$ is so large that using one value to represent the whole region is inaccurate. We therefore split the region into $10\times10$ subregions and calculate $\kappa$ and $Q$ in each subregion. 

\begin{center}
\begin{table*}[htb]
\begin{tabular}{lccccccccc}
Region & \multicolumn{3}{c}{Central} & \multicolumn{3}{c}{Intermediate} & \multicolumn{3}{c}{Outskirt} \\
epoch & 520 & 540 & 560 & 520 & 540 & 560 & 520 & 540 & 560 \\\hline
$0 < \alpha_{\rm full}\leq2$ & 11.7 & 23.4 & 33.2 & 5.70 & 2.30 & 3.01 & 6.86 & 6.91 & 1.49 \\
$\alpha_{\rm full}>2$ & 50.5 & 45.4 & 48.3 & 32.6 & 27.6 & 27.3 & 26.0 & 25.3 & 37.8 \\
$0 > \alpha_{\rm full}\geq-2$ & 4.78 & 8.87 & 6.00 & 15.4 & 12.4 & 7.16 & 17.6 & 16.6 & 1.49 \\
$\alpha_{\rm full}<-2$ & 33.1 & 22.4 & 12.5 & 46.2 & 57.7 & 62.57 & 49.5 & 51.2 & 59.3 \\
\tableline

\end{tabular}
\caption{\label{tab:alpha_w}Percentages of clouds gravitationally bound ($0 > \alpha_{\rm full}\geq-2$) or unbound (all others) according to the full virial parameter $\alpha_{\rm full}$ by region and epoch.}
\end{table*}
\end{center}

\section{Results} \label{sec:results}

\subsection{Virial Parameter Distribution}

Figure~\ref{fig:alpha_hist} and Table \ref{tab:alpha_w}
show the percentages of clouds with potential energy $W \leq 0$ that (1) nevertheless are turbulence-dominated and unbound with magnitude of the full virial parameter $\alpha_{\rm full} < - 2$, (2) are indeed bound with $0 > \alpha_{\rm full} > -2$, and the percentages of gravitationally unbound clouds with $W>0$ that are either (3) dominated by turbulence, with $\alpha_{\rm full}>2$, or (4) dominated by tidal forces with $0 < \alpha_{\rm full} < 2$.  We find significant populations of clouds in all four of these states at all locations and times in our simulation, in agreement with \citet{galeano}. As seen on the right side of the vertical line at $|\alpha_{\rm full}| = 2$ of each panel, gravitationally unbound clouds with $W > 0$ do appear dominated by turbulence throughout epochs 520, 540, and 560. However, the middle panel of each of the three rows shows that gravitationally bound clouds with $W \leq 0$ also {\em appear} dominated by turbulence across all three epochs.

Figure~\ref{fig:signed} compares the classical virial parameter $\alpha_{\rm class}$ calculated using only the self-gravity of the clouds in the spherical approximation to the full virial parameter $\alpha_{\rm full}$ including the complete gravitational potential acting on the cloud from its own self-gravity and the environment. Although there is a weak correlation between $\alpha_{\rm class}$ and $|\alpha_{\rm full}|$, low values of $\alpha_{\rm class} < 2$ are associated with all four types of clouds described above, showing that clouds can be either bound or unbound by tidal forces. Conversely, large, even enormous, values of $\alpha_{\rm class} > 2$ can be associated with bound clouds (those with $0 > \alpha_{\rm full} > -2$). 

A significant fraction of clouds are thus bound because of a compressive external field rather than their own self-gravity. Tidal forces can indeed play a role in unbinding clouds for a majority of clouds that appear to be unbound in a virial analysis.
The fraction of bound clouds depends on both space and time in this simulation.  In the central region, the fraction increases over time, though not monotonically, while it drops by a factor of two in the intermediate region and by an order of magnitude in the outer region. In later sections we examine the properties of each region in more detail.

\subsection{Time Evolution}
Figure~\ref{fig:alpha_hist} shows that the number of clouds in the outskirt region is gradually decreasing over time.
This decrease in the number of clouds may be explained by the completion of cloud collapse and star formation in this region. In contrast, in the intermediate region, there are more clouds at epoch 560 than in 540 or 520, suggesting cloud formation is ongoing there. The same applies to the central region. Note that the central region has around 10--20 times more clouds than the other two regions.

\subsection{Toomre Parameter}

The total Toomre $Q$ value that contains both the contribution of gas and stars defined by Eq. (\ref{eq:qtot}) is shown in Table~\ref{tab:values}. This table reports the quartiles of the distribution of the Q values of the 100 subregions for all three zoom-in regions. There is a clear upward trend towards the central region, where the median value of $Q$ is an order of magnitude higher than in the other two regions. The implied weaker instability seems to contradict the higher percentage of clouds becoming bound over time seen in Table~~\ref{tab:alpha_w}.

However, in the central region there is also a much
larger variation in the distribution of $Q_{\rm gas}$ in different $0.1\times0.1$~kpc subregions than in the other regions. We therefore also measured $Q_{\rm gas}$ in finer subregions of $10\times10$~pc, and found even larger variation. This suggests that (1) gravitationally unstable regions are concentrated on a small volume of the densest gas complex in the central region; and (2) the large $Q$ value of the whole region is dominated by the low-density, volume-filling gas with large gas velocity dispersion created by high shear. The distribution of $Q_{\rm star}$ has a relatively small variation compared to $Q_{\rm gas}$. Therefore, in dense regions $Q$ is dominated by $Q_{\rm gas}$, while in low density regions with high gas velocity dispersion, $Q$ is dominated by $Q_{\rm star}$. In the other two galactic regions, the $Q$ values remain quite similar with much smaller variation across the region. As already noted by \citet{li2006}, $Q > 1$ does not prevent non-axisymmetric collapse, although it does reduce the amount of collapse compared to more unstable regions.

\begin{center}
\begin{table*}

\begin{tabular}{lccccccccc}
Region & \multicolumn{3}{c}{Central} & \multicolumn{3}{c}{Intermediate} & \multicolumn{3}{c}{Outskirt} \\
epoch & 520 & 540 & 560 & 520 & 540 & 560 & 520 & 540 & 560 \\\hline

$Q$  & $30.8_{4.1}^{47.4}$ & $33.9_{1.7}^{43.9}$ & $27.9_{3.3}^{44.7}$ & $4.6_{2.5}^{6.7}$ & $4.3_{2.6}^{6.1}$ & $4.0_{2.5}^{6.1}$ & $1.6_{1.3}^{2.2}$ & $1.6_{1.2}^{2.2}$ & $1.8_{1.2}^{2.4}$ \\
$W^+ / W^-$ & 1.6 & 2.3 & 4.5 & 0.71 & 0.51 & 0.55 & 0.71 & 0.59 & 0.74 \\
(bound / unbound) & 0.050 & 0.097 & 0.064 & 0.183 & 0.142 & 0.077 & 0.214 & 0.199 & 0.015  \\ 
Mean $\Sigma$ [$M_\odot/\mbox{ pc}^2$] & 35.4 & 41.2 & 38.1 & 12.7 & 12.5 & 12.5 & 7.10 & 6.75 & 6.70 \\
\tableline
\end{tabular}

\caption{\label{tab:values} Median and 25--75th percentile of the Toomre parameter $Q$, ratio of number of clouds with positive potential energy $W^+$ to negative $W^-$, ratio of number of clouds with bound virial parameter $0>\alpha_{\rm full}>-2$ to number of unbound clouds ($\alpha_{\rm full} > 0$ or $\alpha_{\rm full} < -2$),  and mean surface density $\Sigma$ for each region and epoch.}
\end{table*}
\end{center}
\subsection{Eigenvalues}

To compare with the tidal forces derived by \cite{ganguly}, the eigenvalues of the tidal tensor of each cloud were computed and plotted against the virial parameter $\alpha_{tt}$ that incorporates both thermal and turbulent energy. Instead of assuming a uniform background potential, analyzing the external tidal field helps us better determine its impact on the formation of both small and large-scale clouds \citep{renaud}. Using the volume-averaged tidal tensor, it is possible to analyze the deformation characteristics of the potential over a specified volume. Because the tensor is symmetric, it can be diagonalized. The eigenvalues $\lambda_{i}$ then represent the strength of the tensor's effect along the direction of each eigenvector.

The sign of the eigenvalue reveals whether that component of the tidal force is compressive or expansive, while their magnitude indicates the its strength. A positive eigenvalue suggests that the gas will expand in the direction associated with that eigenvalue due to the local tidal field, whereas a negative eigenvalue indicates that the gas will compress in that direction \citep{ganguly}.

Taking the ratio of magnitudes of the maximum to the minimum eigenvalue of a symmetric tidal tensor means quantifying the anisotropy of the effect along different directions. To better understand the relative importance of tidal compression versus expansion, we take the ratios of the absolute values of these eigenvalues. This approach helps to separate and analyze the overall strength of tidal effects, regardless of whether they cause compression or expansion \citep{ganguly}. Thus, a relatively high ratio would indicate the tidal forces are much stronger in one direction compared to another while a low ratio implies a fairly uniform strength in all directions. Greater anisotropy suggests that gravity is working more strongly to compress or stretch these structures.

\begin{figure*}
\centering
\includegraphics[width=0.9\linewidth]{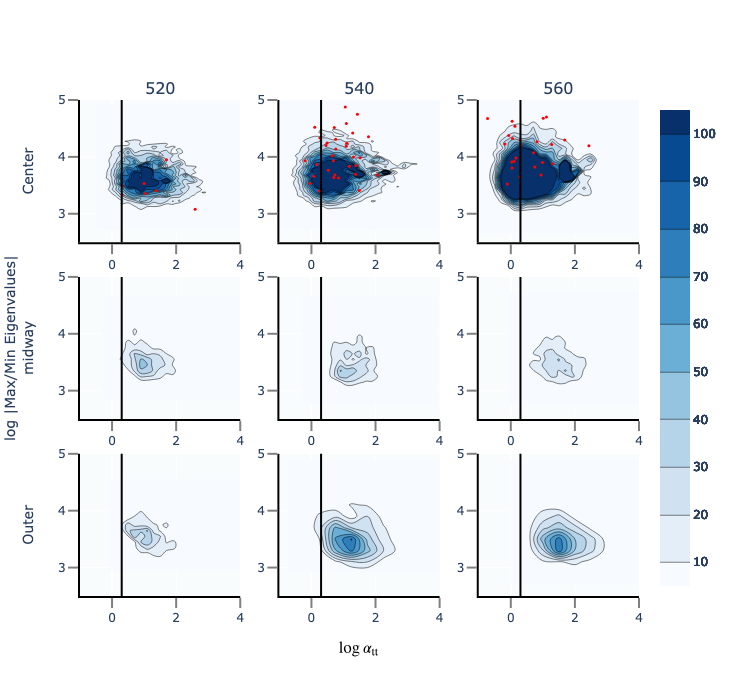}
\caption{\label{fig:eigens} The ratio of the maximum to minimum eigenvalue of the tidal tensor plotted with $\alpha_{tt}$ for the three regions and time epochs (labeled on axes). The red dots indicate fully compressive clouds (all $\lambda < 0$), which only occur in the central region, while not fully compressive clouds are illustrated by the contours. The vertical line in each panel is at $\alpha_{\rm tt} = 2$. These plots were made using a 2D histogram of $50 \times 50$ bins, with the color bar representing the number of clouds per bin.} 
\end{figure*}

As seen in Figure \ref{fig:eigens}, most of the clouds in these epochs have $\alpha_{tt} > 2$. The difference from Figure~\ref{fig:alpha_hist} can be explained by the fact that the definition of $\alpha_{tt}$ incorporates internal energies of clouds that are not included in the full virial parameter $\alpha_{\rm full}$ assessed by \cite{galeano}. 

Only clouds in the central region show tidal forces that are fully compressive, as shown in Figure~\ref{fig:eigens} by red dots, while the other two regions only contain clouds that feel partly or fully extensive tidal forces. In contrast, \cite{ganguly} found most of their clouds to be experiencing tidal compression. 
 
The most likely explanation for this is the different scale of the simulations. \cite{ganguly} used clouds found in the SILCC-Zoom simulations \citep{silcczoom} with 0.06 pc resolution, slightly higher compared with the 0.1--0.2 pc resolution in our case. As a result of the higher resolution, the clouds in SILCC-Zoom can collapse to higher density, and thus are more likely to be identified in regions of tidal compression. They may represent the ultimate fate of the densest clouds in our simulations. Another possibility is that SILCC-Zoom simulations performed the zoom-in by extracting the clouds from the global simulations in periodic boxes, while in our case, the zoom-in regions are simulated together with the galactic background so that the global galactic environments are taken into account simultaneously, providing stronger extensive forces.

\subsection{Heyer Relation}

\begin{figure*}
\centering
\includegraphics[width=0.9\linewidth]{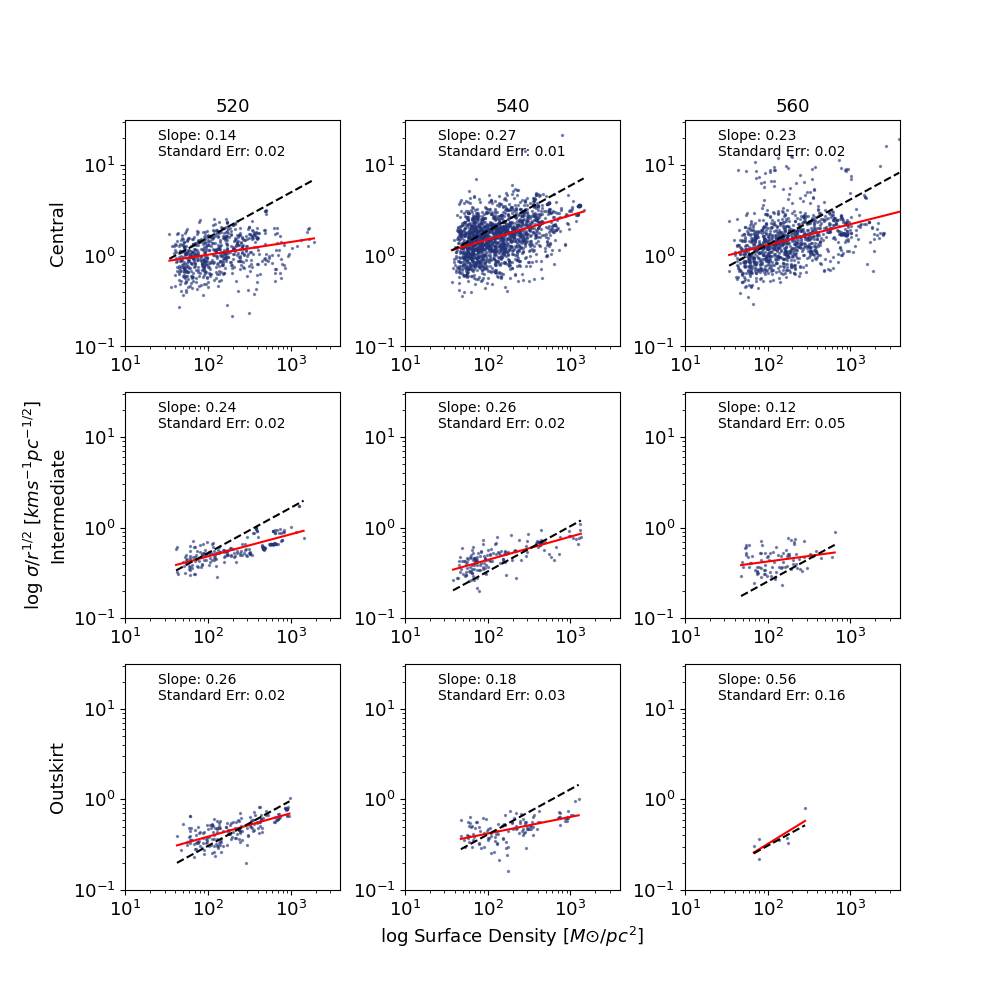}
\caption{\label{fig:heyer} The \citet{heyer} relation for individual clouds at epochs 520, 540, and 560 in the three regions (given in axis labels), using a density threshold to identify clouds of $n>100$~cm$^{-3}$. The solid red line is a linear regression to the data, while the black dashed line follows the virial line. The slope and standard error of the fit are given in each panel, indicating the power-law index of the fit. The linear regression is calculated via the Python library {\sc scipy} \citep{scipy} using the scatterplot (assuming residual normality).}

\end{figure*} 

We have examined the Heyer relation between cloud surface density $\Sigma$ and virial parameter $\sigma / r^{1/2}$, where $\sigma$ is the cloud velocity dispersion and $r$ its radius. The results for each of the regions over time are shown in Figure~\ref{fig:heyer}.

As suggested by \citet{javier}, given the surface density of clouds that ranges from 10--10$^{4} M_{\odot} \mbox{ pc}^{-2}$, the ratio $\sigma / r^{1/2}$ of massive cores spans 0.1--10~km~s$^{-1}\mbox{ pc}^{-1/2}$. For results suggested by \citet{ganguly}, clouds with 10-- 10$^{3} M_{\odot} \mbox{ pc}^{-2}$ have a ratio $\sigma / r^{1/2}$ that spans from 0.1--1~km~s$^{-1}\mbox{ pc}^{-1/2}$. In this figure, more dense and massive clouds exhibit a relatively larger velocity dispersion for their small sizes, which is inferred from the positive slope of the trend line in all cases. Moreover, the data points in Figure~\ref{fig:heyer} are well within the range given by \citet{javier} and \citet{ganguly}.

The lower than expected slopes are likely due to the limited numerical resolution within individual clouds. This suppresses small-scale regions of collapse where large velocity dispersions can appear at high column densities \citep[e.g. see][]{ibanez-mejia2016}.
  
While none of the regions in these models has a cloud population that follow the virial line, they are well within the range of surface densities modeled in \cite{ganguly}. This remains true despite restricting the analysis to clouds above a density threshold of 100~cm$^{-3}$ in this figure. A great majority of clouds in the central region are located below the virial line, meaning the gravitational potential energy overcomes the kinetic energy in these clouds, and they are gravitationally bound.

\section{Summary and Conclusions} \label{sec:conclusions}

We have studied the importance of tidal forces on the evolution of molecular clouds to try to resolve the tension between \citet{galeano}, who find it to be important in both binding and unbinding clouds, and \citet{ganguly}, who find that, although tidal forcing is strongly anisotropic, it only deforms existing clouds. For our study we examined clouds drawn from zoom-in regions in the central region, an intermediate region at roughly the Solar circle equivalent, and the outskirts of a global galaxy model run in AREPO with a novel zoom-in technique that allows target star-forming regions to be resolved at sub-parsec scale. In each region, we analyzed three epochs spaced 2 Myr apart and identified clouds with a dendrogram algorithm. In our analysis, we determined the full gravitational potential energy acting on the clouds, including external tidal terms and used that to compute extended virial parameters as defined by \cite{galeano} and \cite{ganguly}, as well as the local Toomre parameter, the Heyer relation, and the anisotropy of the tidal forcing as measured by the ratio of maximum to minimum eigenvalues of the tidal tensor. Our main conclusions are as follows:
\begin{itemize}
    \item Clouds that appear unbound according to the classical virial parameter may actually be bound according to the full potential energy of the cloud and its environment as seen in Figures~\ref{fig:alpha_hist} and~\ref{fig:signed}. The converse also occurs. Thus, we must take into account the full potential energy of the environment, not just the self-gravity of the cloud itself, to determine the dynamical state of molecular clouds.
    \vspace{1em}
    \item Our results are consistent with \citet{galeano}, in contrast to the claim of \citet{ganguly} that tidal forces are not important in molecular cloud formation. While multiple factors contribute to this discrepancy, 
    the most likely one appears to be the higher densities, smaller sizes, and thus deeper potential wells of the SILCC-Zoom clouds analyzed by \citet{ganguly}. Tidal forces appear to be more important at early times in cloud evolution.
    \vspace{1em}
    \item Based on a preprint version of \citet{ganguly}, \citet{galeano} speculated that the reason for \citet{ganguly} finding that tidal forces were unimportant was that they used average values of the tidal forces across each cloud. However, the final refereed version of \citet{ganguly} clarified that they used the same equation and procedure as \cite{galeano} to compute tidal forces. Our analysis of a different simulation using the same techniques yields results that agree better with \citet{galeano}, consistent with our simulations also being global models. Thus, it appears more likely that the different conclusions come from the nature of the different simulations analyzed in the two papers rather than any substantive difference in methodology. 
    \vspace{1em}
    \item The only region in which fully compressive tidal forces occur on any clouds in our model is the center (Fig.~\ref{fig:eigens}).  Most clouds have at least one extensive axis, including all clouds in the intermediate and outskirts regions.
    \vspace{1em}
    \item We find that the clouds identified by our analysis do not quantitatively reproduce the observed Heyer relation (Fig. \ref{fig:heyer}), although they do qualitatively show a positive slope of the relation at all times in the central and intermediate regions in agreement with observations.The failure to reproduce the relation may well be due to the comparatively low resolution of the internal structure of individual clouds, which will suppress increasing velocity dispersions produced by collapse at small scales.
    \vspace{1em}
    \item Most of the central region has a higher median value of the Toomre instability parameter $Q$ than other regions, due to the high shear in the region, but the dense clouds occupying a small fraction of the volume have far lower values. This explains why more clouds in this region become bound during the simulation despite the high median value. It should be noted that all regions at all epochs do satisfy the azimuthal Toomre gravitational stability criterion. However, non-azimuthal gravitational instability can still efficiently assemble dense gas to form bound clouds.
    
\end{itemize}

\begin{acknowledgments}
We thank J. Ballesteros-Paredes for useful discussions, and the anonymous referee for detailed reports pointing out an error in our analysis of the thermal pressure as well as leading to substantial improvements in our presentation.
HL is supported by the National Key R\&D program of China No.\ 2023YFB3002502
and the National Natural Science Foundation of China under No.\ 12373006.
M-MML was partly supported by US NSF grant AST23-07950 and by NSF grant PHY23-09135 to the Kavli Institute for Theoretical Physics and
acknowledges Interstellar Institute's program ``II6'' and the
Paris-Saclay University's Institut Pascal for hosting discussions that
nourished the development of some of the ideas behind this work.
\end{acknowledgments}

\bibliographystyle{aasjournal}

\bibliography{ms}

\end{document}